\documentclass[aps,prl,reprint,superscriptaddress,floatfix]{revtex4-1}
\usepackage{amsmath}
\usepackage{amsfonts}
\usepackage{graphicx}
\usepackage[colorlinks,allcolors=blue]{hyperref}

\newcommand{\dpar}[1]{\left(#1\right)}
\newcommand{\dsqr}[1]{\left[#1\right]}

\renewcommand{\P}{\mathbb{P}}

\DeclareMathOperator{\var}{Var}
\DeclareMathOperator{\tr}{tr}

\newcommand{\beq}{\begin{equation}}
\newcommand{\eeq}{\end{equation}}

\begin{document}
\title{Paradoxical probabilistic behavior for strongly correlated many-body classical systems} 
\author{Max Jauregui}
\email{jauregui@cbpf.br}
\affiliation{Centro Brasileiro de Pesquisas F\'isicas and National Institute of Science and Technology for Complex Systems, Rua Xavier Sigaud 150, Rio de Janeiro 22290-180, RJ, Brazil}
\author{Constantino Tsallis}
\email{tsallis@cbpf.br}
\affiliation{Centro Brasileiro de Pesquisas F\'isicas and National Institute of Science and Technology for Complex Systems, Rua Xavier Sigaud 150, Rio de Janeiro 22290-180, RJ, Brazil}
\affiliation{Santa Fe Institute, 1399 Hyde Park Road, Santa Fe, NM 87501, USA}

\begin{abstract}
Using a simple probabilistic model, we illustrate that a small part of a strongly correlated many-body classical system  can show a paradoxical behavior, namely asymptotic stochastic independence.  We consider a triangular array such that each row is a list of $n$ strongly correlated random variables. The correlations are preserved even when $n\to\infty$, since the standard central limit theorem does not hold for this array. We show that, if we choose a fixed number $m<n$ of random variables of the $n$th row and trace over the other $n-m$ variables, and then consider $n\to\infty$, the $m$ chosen ones can, paradoxically, turn out to be independent. However, the scenario can be different if $m$ increases with $n$. Finally, we suggest a possible experimental verification of our results near criticality of a second-order phase transition.\end{abstract}

\maketitle

As well known, sensible differences exist between quantum and classical correlations. The basic reason for that is that the quantum description of a many-body system allows the appearance of entangled states which do not have a clear classical counterpart. For instance, a study of   a finite-length part of an infinitely long quantum spin chain has  shown, among other things, that quantum correlations remain present in the subchain \cite{CarusoTsallis2008}. The correlations are strong enough to mandate the replacement of the additive von Neumann entropy for thermodynamical purposes. Indeed, that entropy is known to be nonextensive for such systems, whereas the nonadditive $q$-entropy \cite{Tsallis1988} for a special value of the index $q$ re-establishes the desirable thermodynamical extensivity. For a quantum critical phenomenon of a one-dimensional many-body model belonging to the universality class associated with the central charge $c$, the value of that index is given by
\beq
q=\frac{\sqrt{9+c^2}-3}{c} \,.
\eeq
For instance, for the first-neighbour Ising ferromagnet in the presence of a transverse magnetic field, we have $c=1/2$, hence $q=\sqrt{37}-6 \simeq 0.08$ .

The $q$-entropy associated with a density matrix $\rho$ is defined by
\beq
S_q(\rho):=k\frac{1-\tr \rho^q}{q-1}\,,
\eeq
for any real $q\ne 1$ ($S_1$ is the von Neumann entropy $-k\tr (\rho \ln \rho)$). This entropy is at the core of the so-called nonextensive statistical mechanics \cite{Tsallis1988,Gell-MannTsallis2004,Tsallis2009}, which focuses on complex systems such as those involving long-range interactions or other sources of strong correlations (typically causing an ergodicity breakdown). Extremization of the $q$-entropy with appropriate constraints leads to the so-called {\it $q$-exponential} and {\it $q$-Gaussian distributions}. The $q$-exponential function is defined by
\beq
\label{qexp}
e_q^x:=
\begin{cases}
[1-(1-q)x]^{1/(1-q)}&\text{for any real $q\ne 1$}\\
e^x&\text{for $q=1$}
\end{cases}
\eeq
for any $x$ such that $1+(1-q)x>0$. The $q$-Gaussian distribution with real parameters $q<3$ and $\beta>0$ is characterized by the density
\beq
\label{qG}
G_q(\beta,x):=
\begin{cases}
\frac{\sqrt{\beta}}{N_q}e_q^{-\beta x^2}&\text{if $1+(q-1)\beta x^2>0$}\\
0&\text{otherwise,}
\end{cases}
\eeq
where
\beq
N_q:=
\begin{cases}
\frac{2^{\frac{3-q}{1-q}}[\Gamma(\frac{2-q}{1-q})]^2}{\sqrt{1-q}\Gamma(\frac{2(2-q)}{1-q})}&\text{for $q<1$}\\
\sqrt{\pi}&\text{for $q=1$}\\
\frac{\sqrt{\pi}\Gamma(\frac{3-q}{2(q-1)})}{\sqrt{q-1}\Gamma(\frac{1}{q-1})}&\text{if $1<q<3$.}
\end{cases}
\eeq
From~(\ref{qG}), we notice immediately that a $q$-Gaussian distribution has compact support whenever $q<1$. If $1\le q<3$, the support of a $q$-Gaussian distribution is the whole real line. By the way, let us mention that, if $\beta <0$, $q$-Gaussian distributions are normalizable for $q \ge 3$.

Both the $q$-exponential and $q$-Gaussian distributions appear in a large number of natural, artificial and social systems, e.g., in long-range-interacting many-body classical Hamiltonian systems \cite{PluchinoRapisardaTsallis2007,CirtoAssisTsallis2014,ChristodoulidiTsallisBountis2014}, cold atoms in dissipative optical lattices \cite{DouglasBergaminiRenzoni2006,LutzRenzoni2013}, dusty plasmas \cite{LiuGoree2008}, in the study of the over-damped motion of interacting particles \cite{AndradeSilvaMoreiraNobreCurado2010,RibeiroNobreCurado2012a,RibeiroNobreCurado2012b}, in high energy physics \cite{WongWilk2012,WongWilk2013a,WongWilk2013b,CirtoTsallisWongWilk2014} and in biology \cite{UpadhyayaRieuGlazierSawada2001}.

A subsystem of a correlated system is in principle expected to also be correlated, even in classical systems. For instance, let us consider the following triangular array of random variables which take the values $0$ or $1$:
\beq
\label{array}
\begin{array}{ccccc}
X_{1,1}&&&&\\
X_{2,1}&X_{2,2}&&&\\
\vdots&\vdots&\ddots&&\\
X_{n,1}&X_{n,2}&\cdots&X_{n,n}&\\
\vdots&\vdots&&\vdots&\ddots
\end{array}
\eeq
Each row can be thought as an outcome of the experiment of tossing $n$ coins, where $X_{n,i}=1$ if the $i$th coin falls head and $X_{n,i}=0$ otherwise. For the distribution of the $n$th row of (\ref{array}) we will use one introduced by Rodriguez \textit{et al} \cite{RodriguezSchwammleTsallis2008}. More precisely, fixed any list $(x_1,\dots,x_n)$ formed with the digits $0$ and $1$ such that $x_1+\cdots+x_n=k$, we define
\beq
\label{joint}
\P[X_{n,1}=x_1,\dots,X_{n,n}=x_n]:=\frac{1}{Z_{q,n}}\binom{n}{k}^{-1}e_q^{-t_{q,n,k}^2}\,,
\eeq
where $q<1$ is a real parameter,
\beq
t_{q,n,k}:=\frac{1}{\sqrt{1-q}}\dsqr{1-2\dpar{\frac{k+1}{n+2}}}
\eeq
and
\beq
Z_{q,n}:=\sum_{k=0}^ne_q^{-t_{q,n,k}^2}\,.
\eeq

\begin{figure}[t]
\centering
\includegraphics[width=0.4\textwidth,keepaspectratio]{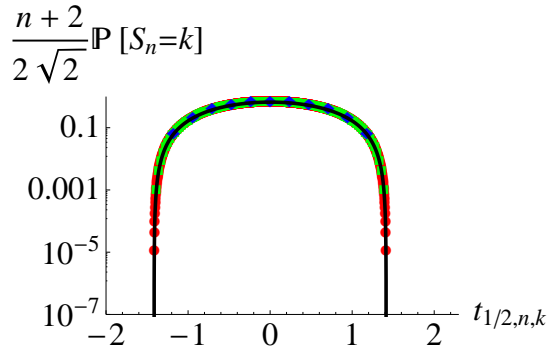}
\caption{Representation of $\frac{(n+2)}{2\sqrt{2}}\P[S_n=k]$ as a function of $t_{1/2,n,k}$ for $n=10$, $100$ and $1000$, where we have considered $q=1/2$. The solid curve represent the right hand side of~(\ref{approxm}).}
\label{approxm.fig}
\end{figure}

After some algebra, it follows from~(\ref{joint}) that
\beq
\label{bern}
\P[X_{n,i}=1]=\P[X_{n,i}=0]=\frac{1}{2}\,.
\eeq
This implies that the random variables in each row of~(\ref{array}) are not independent since the right hand side of~(\ref{joint}) is {\it not} equal to $1/2^n$. Moreover, putting $S_n:=X_{n,1}+\cdots+X_{n,n}$, it can be verified that
\beq
\label{approxm}
\frac{(n+2)\sqrt{1-q}}{2}\P[S_n=k]\sim G_q(1,t_{q,n,k})
\eeq
as $n\to\infty$ (see figure \ref{approxm.fig}). It becomes now transparent why we have adopted, for our illustration, the distribution in~(\ref{joint}).

\begin{figure}[t]
\centering
\includegraphics[width=0.4\textwidth,keepaspectratio]{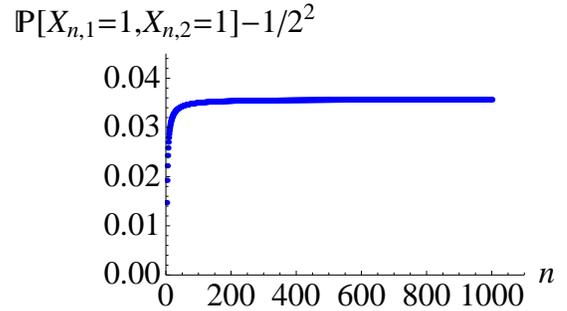}
\caption{Representation of $\P[X_{n,1}=1,X_{n,2}=1]-1/2^2$ as a function of $n$ for $q=1/2$. We see that the the curve goes away from $0$ as $n$ increases. This numerical result is confirmed by the $q=0$ case, which turns out to be analytically tractable, and provides  $\P[X_{n,1}=1,X_{n,2}=1]-1/2^2=1/20 \,, \forall n$. The present available results are consistent with a value of  $\lim_{n\to\infty}  \Bigl\{  \P[X_{n,1}=1,X_{n,2}=1]-1/2^2 \Bigr\}  $ which approaches zero for $q$ approaching one from below.}
\label{pairm}
\end{figure}

If we analyze the marginal probabilities of the first two random variables of the $n$th row of (\ref{array}), we will observe that, analogously to the case of two spins in a quantum spin chain \cite{CarusoTsallis2008}, these two random variables are correlated (see figure \ref{pairm}), which is something we may intuitively expect. However, as we will see later on, this property does not necessarily follow from the fact that the random variables in each row of (\ref{array}) are strongly correlated.

Suppose now that we have the array (\ref{array}) and the distribution given in (\ref{joint}), but this time with $q\ge 1$ and
\beq
\label{tk}
t_{q,n,k}:=\sqrt{n+1}\dpar{\frac{k+1}{n+2}-\frac{1}{2}}\,.
\eeq
It follows immediately from~(\ref{joint}) that, as in the $q<1$ case,
\begin{multline}
\label{ex}
\P[X_{n,\pi(1)}=x_1,\dots,X_{n,\pi(n)}=x_n]=\\
\P[X_{n,1}=x_1,\dots,X_{n,n}=x_n]
\end{multline}
for any permutation $\pi$ of $\{1,\dots,n\}$. This allows us to use the reduced notation
\beq
r_{q,n,k}:=\P[X_{n,1}=x_1,\dots,X_{n,n}=x_n]\,,
\eeq
where $(x_1,\dots,x_n)$ is any list formed with the digits $0$ and $1$ such that and $x_1+\cdots+x_n=k$. Moreover, it follows from (\ref{ex}) that
\beq
s_{q,n,k}:=\P[S_n=k]=\binom{n}{k}r_{q,n,k}\,.
\eeq

Let us incidentally mention that, for $q>1$, it has been  shown \cite{RuizTsallis2012,RuizTsallis2013} that the corresponding large deviation theory involves $q$-exponentials instead of the standard exponentials. More precisely, it has been shown numerically that, as $n$ increases, the probability of large deviations $\P[S_n\le nx]$, $x<1/2$, decay to zero like a $q'$-exponential, where $q'>1$ is some function $q^\prime (q)$ such that $q^\prime(1)=1$, thus recovering the usual theory. 

Also, it follows from~(\ref{tk}) that $t_{q,n,n-k}=-t_{q,n,k}$ and, consequently, $r_{q,n,n-k}=r_{q,n,k}$. This implies that $\P[S_n\le nx]=\P[S_n\ge n(1-x)]$ for every $x<1/2$. The fact that these probabilities converge to zero as $n\to\infty$ means that the weak law of large numbers holds for~(\ref{array}). It is easy to see from~(\ref{joint}) that $r_{q,n,k}\ne r_{q,n+1,k}+r_{q,n+1,k+1}$. Nevertheless, it has been verified \cite{RodriguezSchwammleTsallis2008} that the equality ({\it Leibniz triangle rule}) holds asymptotically when $n\to\infty$, i.e., $\lim_{n\to\infty} \frac{r_{q,n+1,k}+r_{q,n+1,k+1}}{r_{q,n,k}}=1$. 

\begin{figure}[t]
\centering
\includegraphics[width=0.4\textwidth,keepaspectratio]{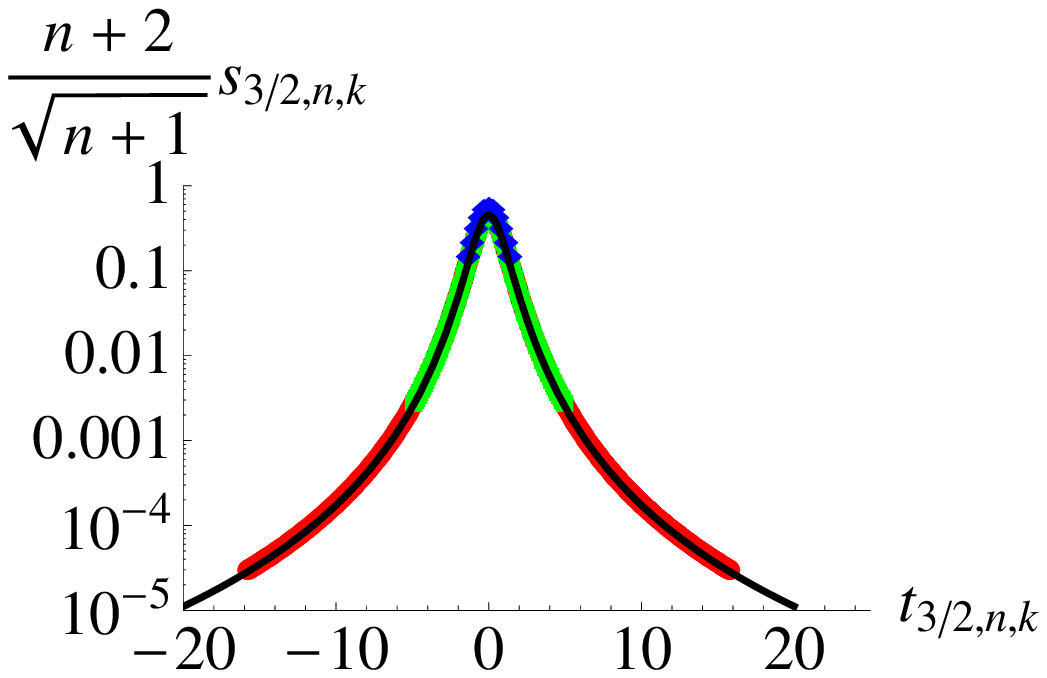}\quad\includegraphics[width=0.4\textwidth,keepaspectratio]{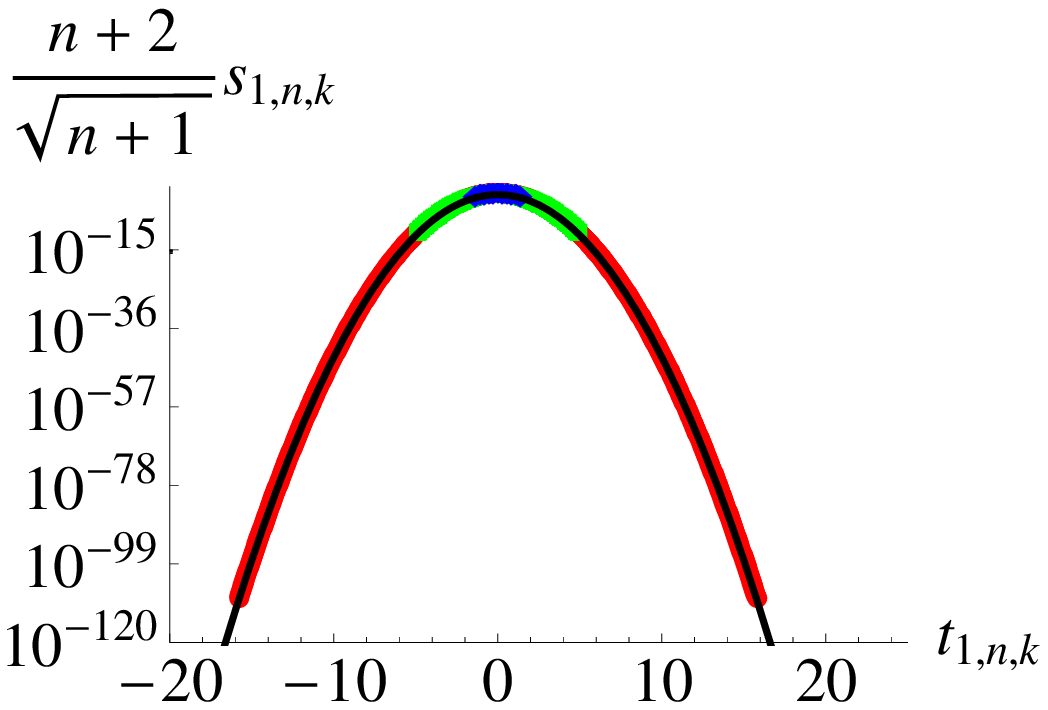}
\caption{Representation of $\frac{n+2}{\sqrt{n+1}}s_{q,n,k}$ as a function of $t_{q,n,k}$ for $n=10$, $100$ and $1000$, where we have considered $q=3/2$ (top) and $q=1$ (bottom). In both cases, the solid curves represent the right hand side of~(\ref{approx}).}
\label{approx.fig}
\end{figure}

Also for $q\ge 1$, (\ref{bern}) holds, meaning that the random variables in each row of~(\ref{array}) are not independent. Moreover, the distribution of $S_n$, after appropriate scaling, approximates, as $n$ increases, to a $q$-Gaussian distribution with parameters $q\ge 1$ and $\beta=1$ \cite{RuizTsallis2012}. More precisely,
\beq
\label{approx}
\frac{n+2}{\sqrt{n+1}}s_{q,n,k}\sim G_q(1,t_{q,n,k})
\eeq
for large values of $n$ (see figure \ref{approx.fig}).

For $n>1$, let us focus on the analysis of $m<n$ random variables of the $n$th row of~(\ref{array}). Given any list $(x_1,\dots,x_m)$ formed with the numbers $0$ and $1$ such that $x_1+\cdots+x_m=l$, we obtain from~(\ref{joint}) that the marginal probabilities are given by
\begin{multline}
\label{marg}
\P[X_{n,1}=x_1,\dots,X_{n,m}=x_m]=\\
\frac{1}{Z_{q,n}}\sum_{j=0}^{n-m}\binom{n-m}{j}\binom{n}{j+l}^{-1}e_q^{-t_{q,n,j+l}^2}\,.
\end{multline}
By virtue of~(\ref{ex}), the joint distribution of any subset containing $m$ random variables of the $n$th row of~(\ref{array}) is given by the right hand side of~(\ref{marg}).
In order to simplify the notation, we will use
\beq
p_{q,n,m}:=\P[X_{n,1}=1,\dots,X_{n,m}=1]\,.
\eeq

\begin{figure}[t]
\centering
\includegraphics[width=0.4\textwidth,keepaspectratio]{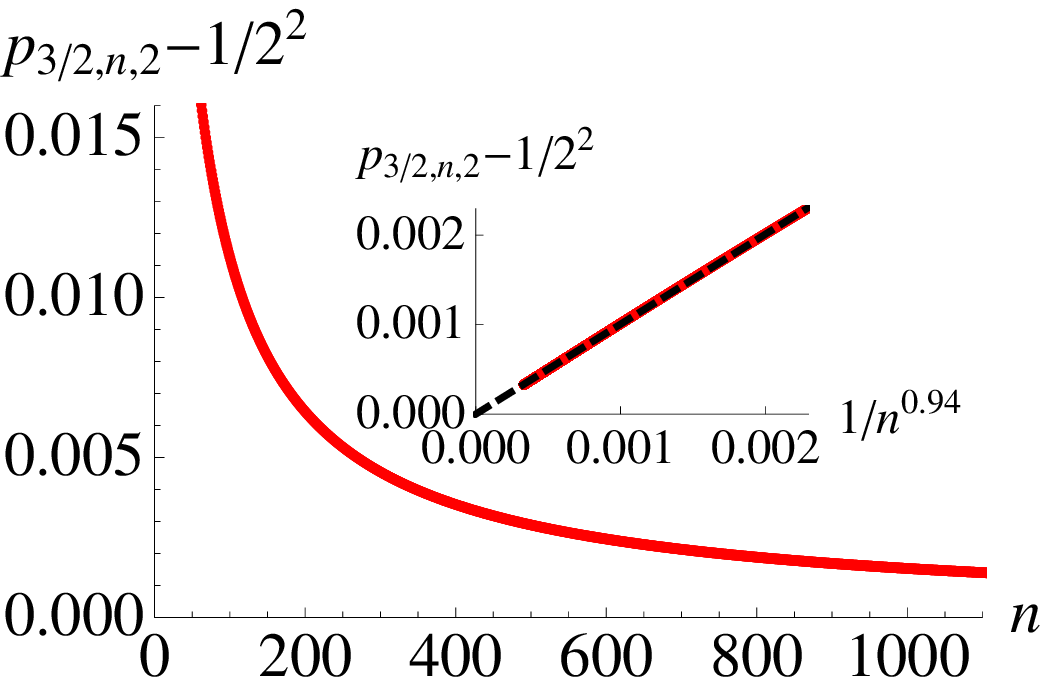}\quad\includegraphics[width=0.4\textwidth,keepaspectratio]{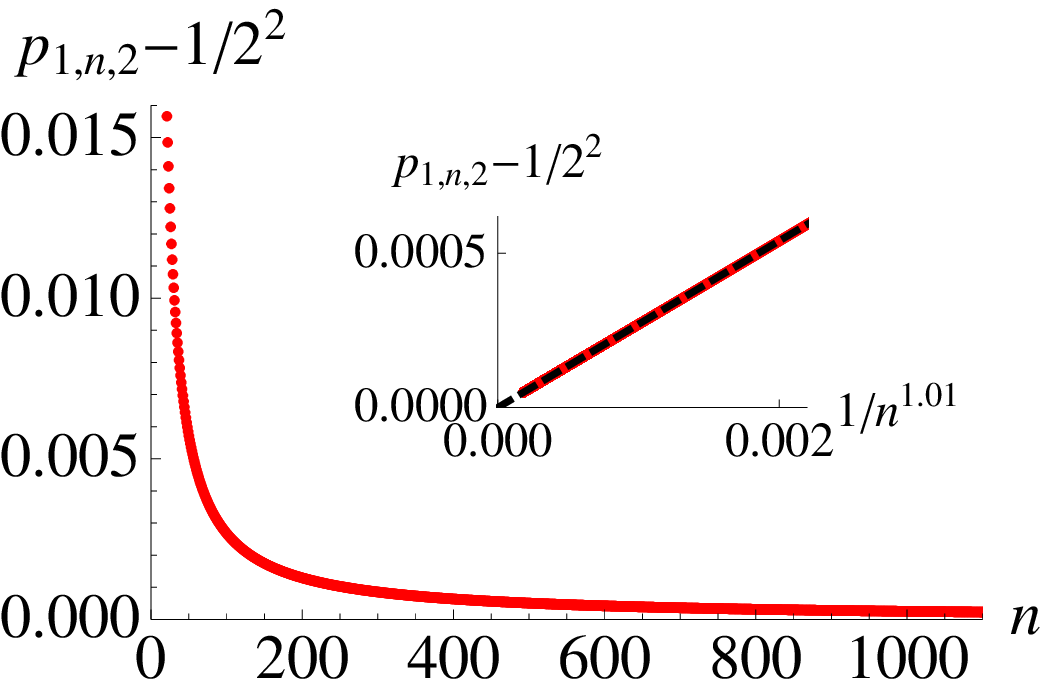}
\caption{Representation of $p_{q,n,2}-1/2^2$ as a function of $n$, where we have considered $q=3/2$ (top) and $q=1$ (bottom). In both cases we notice that $p_{q,n,2}$ approaches $1/2^2$ as $n$ increases. The insets suggest that $p_{q,n,2}-1/2^2$ decay to zero like a power law in both cases. The exponent of $n$ appears to slightly depend on $q$: see also figure \ref{others.fig}.}
\label{pair.fig}
\end{figure}

We numerically verify that $p_{q,n,2}$ approaches $1/2^2$ as $n$ increases (see figure~\ref{pair.fig}). This implies that any two variables  of rows of~(\ref{array}) that are sufficiently far below are, surprisingly enough, asymptotically independent. This appears to be, in the present case, the reason why the law of large numbers holds for~(\ref{array}). Indeed, pairwise independence is sufficient for this law to emerge \cite{Durrett}. Moreover, fixing $m>0$, it can be verified that
\begin{multline}
\var(X_{n,1}+\cdots+X_{n,m})=\\
m\dpar{\frac{1}{2}-p_{q,n,2}}+m^2\dpar{p_{q,n,2}-\frac{1}{4}}\,,
\end{multline}
which, as $n$ increases, approaches the value corresponding to independence, namely $m/2$ .
\begin{figure}[ht]
\centering
\includegraphics[width=0.4\textwidth,keepaspectratio]{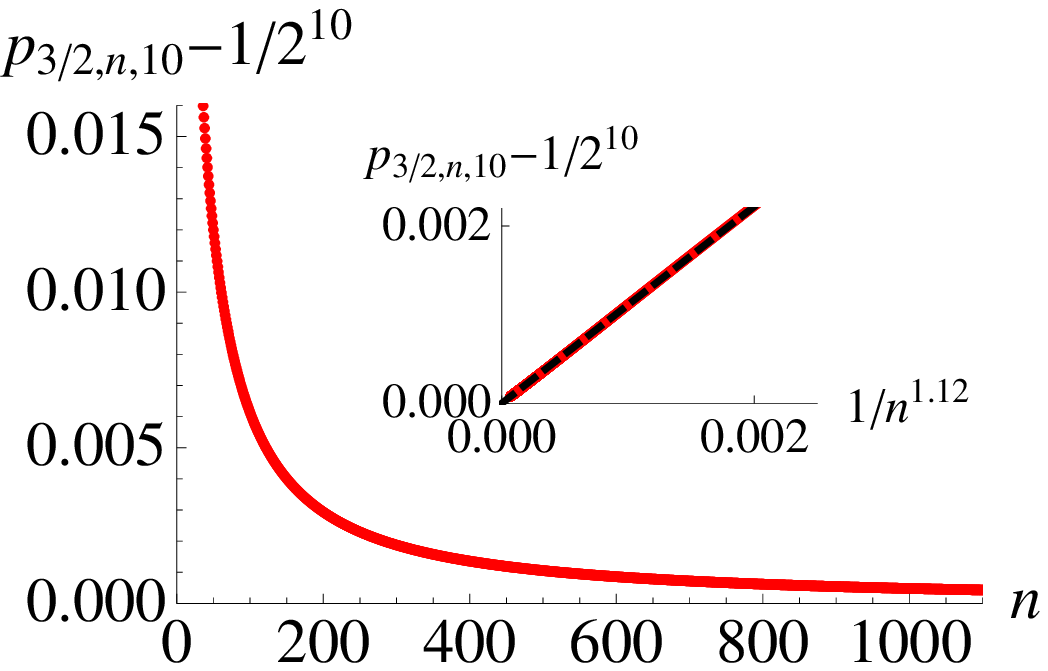}\quad\includegraphics[width=0.4\textwidth,keepaspectratio]{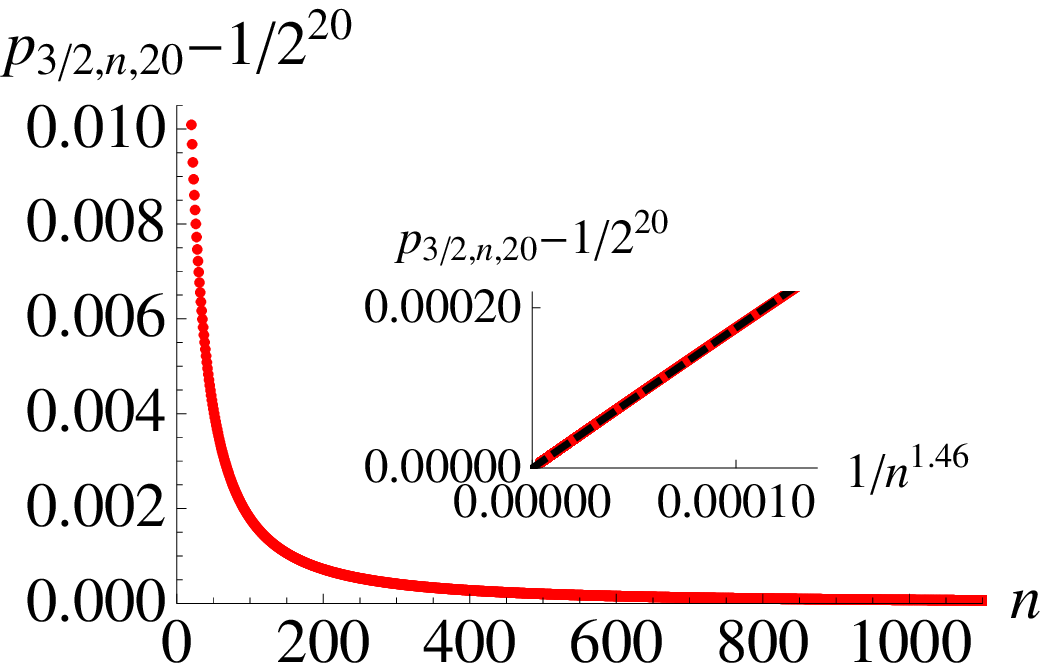}
\caption{Representation of $p_{3/2,n,9}-1/2^9$ (top) and $p_{3/2,n,20}-1/2^{20}$ (bottom) as functions of $n$. The insets suggest that both functions decay to zero like a power law. The exponent of $n$ appears to depend on $m$ for a given value of $q$.}
\label{others.fig}
\end{figure}

\begin{figure}[t]
\centering
\includegraphics[width=0.4\textwidth,keepaspectratio]{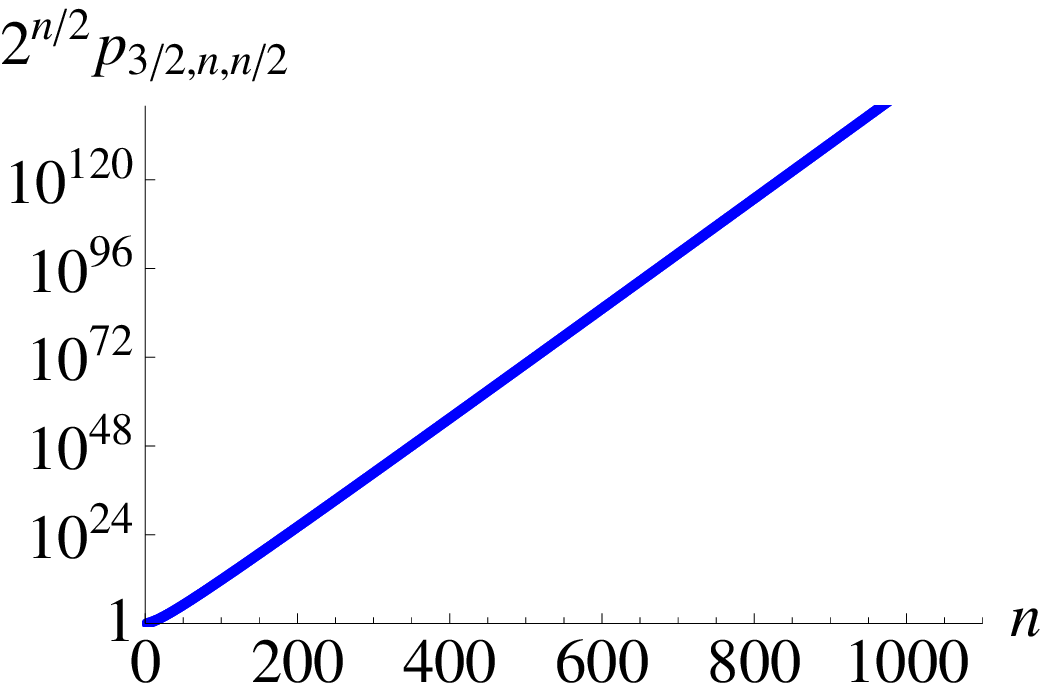}\quad\includegraphics[width=0.4\textwidth,keepaspectratio]{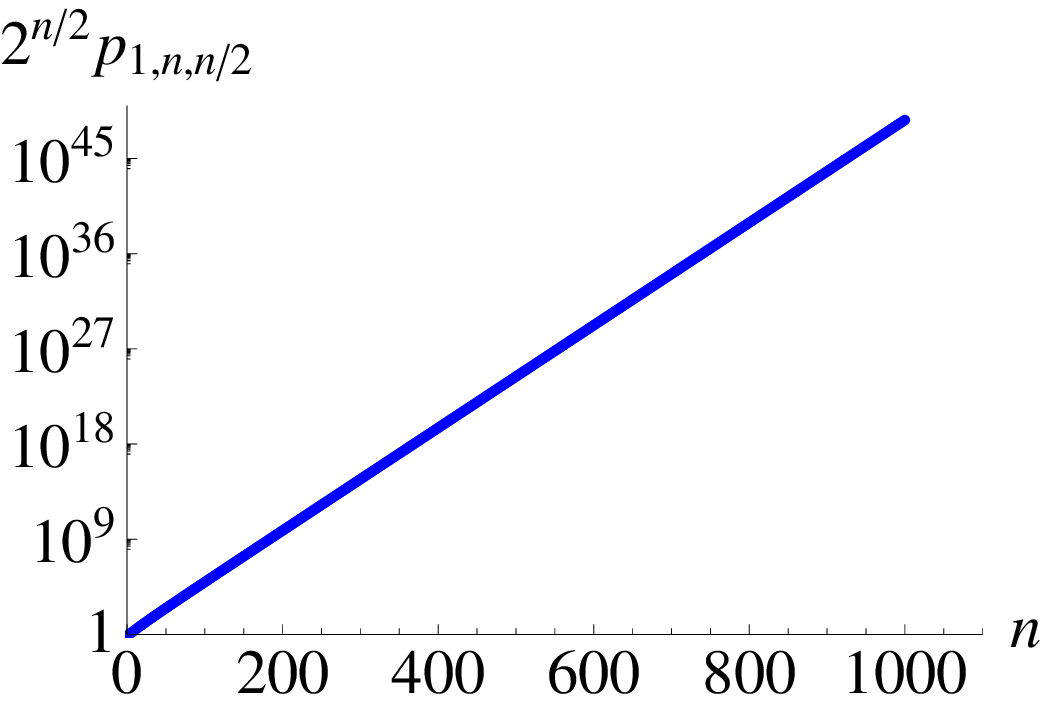}
\caption{Representation of the ratio of $p_{q,n,n/2}$ and $1/2^{n/2}$ as a function of $n$ for $q=3/2$ (top) and $q=1$ (bottom). Clearly, this ratio is increasingly different from $1$ when $n$ is very large. To construct this graph, we have considered even values of $n$.}
\label{half}
\end{figure}
\begin{figure}[t]
\centering
\includegraphics[width=0.4\textwidth,keepaspectratio]{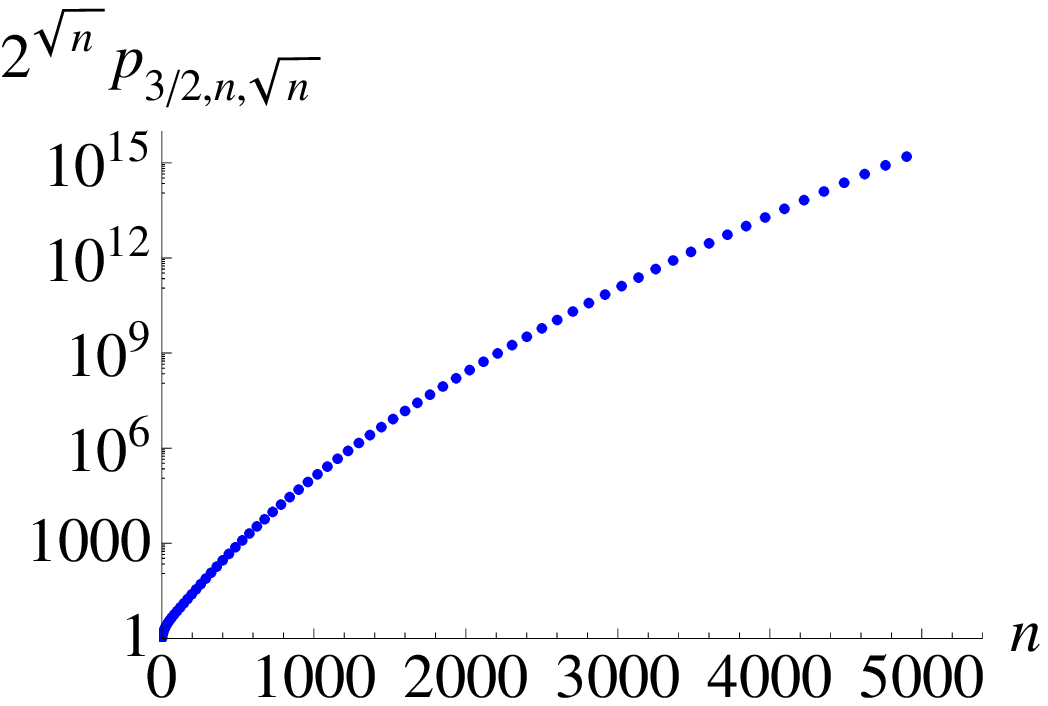}\quad\includegraphics[width=0.4\textwidth,keepaspectratio]{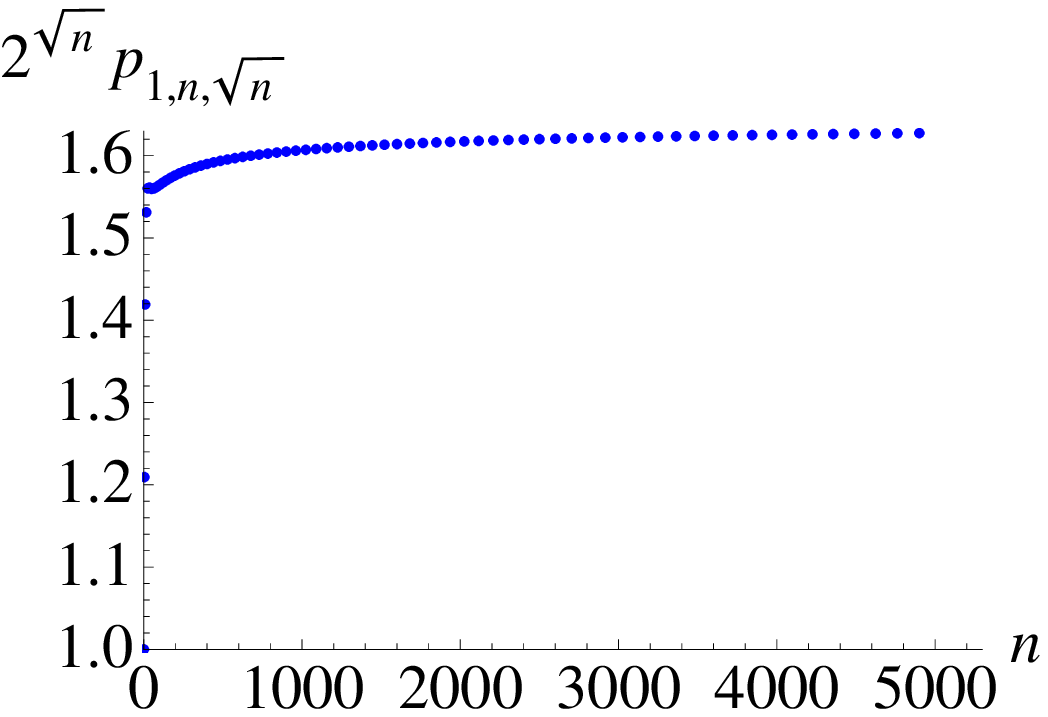}
\caption{Representation of the ratio of $p_{q,n,\sqrt{n}}$ and $1/2^{\sqrt{n}}$ as a function of $n$ for $q=3/2$ (top) and $q=1$ (bottom). Clearly, this ratio is different from $1$ when $n$ is very large. To construct this graph, we have considered perfect squares for $n$.}
\label{sqrt}
\end{figure}

If instead of choosing a pair of random variables, we choose $m>2$ ones from the $n$th row of~(\ref{array}), we will obtain that $p_{q,n,m}$ approaches $1/2^m$ as $n$ increases, as illustrated in  figure~\ref{others.fig}. This implies, like in the case $m=2$, that the correlations among $m$ random variables of a row of~(\ref{array}) asymptotically vanish if $n\to\infty$.  However, the scenario seems to be different if we choose $m_n$ random variables of the $n$th row of~(\ref{array}), where $m_1,m_2,\dots$ is some increasing sequence of positive integers with $m_n\le n$ (see figures \ref{half} and \ref{sqrt}). This is something that is not surprising since the $n$th row of~(\ref{array}) does not become independent as $n\to\infty$; otherwise~(\ref{array}) would obey the standard central limit theorem, contradicting~(\ref{approx}).

We conclude that, for $q\ge 1$ (i.e., unbounded support), if we are restricted to analyze a prefixed finite number of random variables in each row of~(\ref{array}), it is, for the present model, impossible to decide whether the chosen random variables are part of an independent or a correlated superset of random variables. However, this ambiguity is removed if we focus on a number of random variables which increases with $n$.

In science, in many circumstances, we want to analyze a growing system. For instance, the evolution of a microbiological culture, the propagation of an epidemic, among others. However, sometimes by technical difficulties, we are just allowed to study parts of the system with a determined size, which is much smaller than the size of the whole system. Naturally, we can think that analyzing several parts of that size can be sufficient to reach a conclusion about the whole system. However, as we have seen, this procedure may lead to the statement of false claims about the entire system, specially if strong correlations are present in it. In contrast, the results obtained by looking at a growing part of the system may be generalized to the whole system.

As a physical illustration of the present paradoxical results, we suggest the following experiment. Consider a macroscopic sample of some material which presents a second-order phase transition. We will perform measurements (e.g., magnetic or electric susceptibility) on a part with fixed size of the large sample. This part can be macroscopic as well but much smaller than the whole sample. As we adjust the temperature of the sample to values near the critical one, the correlation length increases and eventually surpasses the size of the part we are focusing on. At this stage, the microscopic constituents in the whole sample clearly are strongly correlated. However, this correlation might, interestingly enough, {\it not} be detectable in the subsystem that we are studying.

Let us finally emphasize that the present surprising effects possibly are linked to the classical nature of the system, and to the fact that the support of the attractor is unbounded. Indeed, quantum correlations, as well as classical ones with bounded support, are of a different nature, and therefore these correlations do persist  at the level of a part of the system, even if the size of the entire system keeps increasing.

\begin{acknowledgements}
We acknowledge L.~J.~L. Cirto and E.~M.~F. Curado for fruitful discussions. Partial financial support from CNPq and Faperj (Brazilian agencies) are also acknowledged. One of us (CT) has also benefitted from partial financial support from the John Templeton Foundation.
\end{acknowledgements}

\bibliography{articles,books}
\bibliographystyle{apsrev4-1}
\end{document}